\pdfoutput=1
\documentclass{article}
\usepackage{spconf,amsmath,amssymb,graphicx}

\usepackage{booktabs}
\usepackage{multirow}
\usepackage{makecell}
\usepackage{listings}
\usepackage{cite}
\usepackage{hyperref}
\usepackage[table]{xcolor}

\usepackage{microtype}
\usepackage{comment}

\definecolor{best}{RGB}{183, 222, 187}
\definecolor{second}{RGB}{222, 239, 224}
\definecolor{baseline}{RGB}{242, 242, 242}

\newcommand{\bestcell}[1]{\cellcolor{best}#1}
\newcommand{\secondcell}[1]{\cellcolor{second}#1}
\newcommand{\basecell}[1]{\cellcolor{baseline}#1}

\newcommand{\bestshade}[1]{%
  {\setlength{\fboxsep}{1pt}\colorbox{best}{#1}}}
\newcommand{\secondshade}[1]{%
  {\setlength{\fboxsep}{1pt}\colorbox{second}{#1}}}

\newcommand{\parhead}[1]{\vspace{2pt}\noindent\textbf{#1}\ }

\title{AEGIS: Audio Endogenous Guarding via Internal Signals \\Against Large Audio-Language Model Jailbreaks}

\name{Yu-Ling Liao\sthanks{Equal contribution.}, Tzu-Chin Chiu\footnotemark[1], Zong-You Chen\footnotemark[1], Chi-Lei Tsai, Shao-Yuan Lo}
\address{National Taiwan University\\
\texttt{\{r14944077, b11902105, r13922194, b11201048, shaoyuan\}@ntu.edu.tw}}

\begin{document}
\maketitle

\begin{abstract}

Large audio-language models (LALMs) expand language models to process and interpret audio, but also expose them to heterogeneous audio jailbreaks. We ask whether successful jailbreaks reflect failures to recognize harmful intent or failures occurring after such recognition. Layer-wise probing reveals the latter: risk-related information remains decodable from intermediate representations, yet the internal risk signal fails to translate into refusal in later-layer processing. We identify this discrepancy as the \textit{risk-to-refusal gap}. Building on this finding, we propose \textsc{Aegis}, a detect-then-intervene defense whose mid-layer risk gate selectively activates downstream safety adapters. Across six LALMs and three heterogeneous audio jailbreak benchmarks, \textsc{Aegis} reduces the average unsafe rate from 17.9\% to 0.4\%, while causing only a marginal increase in over-refusal on benign inputs. These results establish selective internal intervention as an effective path toward more robust refusal in LALMs. The code is available at \url{https://github.com/azzzzliao/aegis-audio-defense}.

\begin{keywords}
large audio-language models, jailbreaks, speech safety, representation-level intervention
\end{keywords}
\end{abstract}
\section{Introduction}

Recent advances in large audio-language models (LALMs) extend language models with auditory understanding~\cite{Qwen2Audio, gemmateam2026gemma4, fu2025vita, ultravox, liu2025voxtral, microsoft2025phi4minitechnicalreportcompact}. However, the audio modality exposes LALMs to heterogeneous jailbreaks, including semantic and multilingual obfuscation, paralinguistic variation, compositional audio, and adversarial waveform perturbations~\cite{peng2026jalmbenchbenchmarkingjailbreakvulnerabilities, song2025audiojailbreakopencomprehensive, yang2026speechaudiocompositionalattacksmultimodal, kang2025advwave}.

Existing defenses mitigate audio jailbreaks through input-side filtering, safety training, and representation-level intervention~\cite{yang2026speechaudiocompositionalattacksmultimodal, wang2026omnisafetycrossmodalityconflictvulnerabilities, yang2025reshapingrepresentationspacebalance, lin2026sarsteersafeguardinglargeaudiolanguage}. 
Although these methods can reduce unsafe responses at different stages of generation, they do not reveal where the model's safety alignment breaks down when a jailbreak succeeds.
In particular, \textit{does a successful jailbreak arise because the model fails to recognize harmful intent, or because recognized risk fails to trigger refusal?}

\begin{figure}[t]
    \centering
        \includegraphics[width=\linewidth]{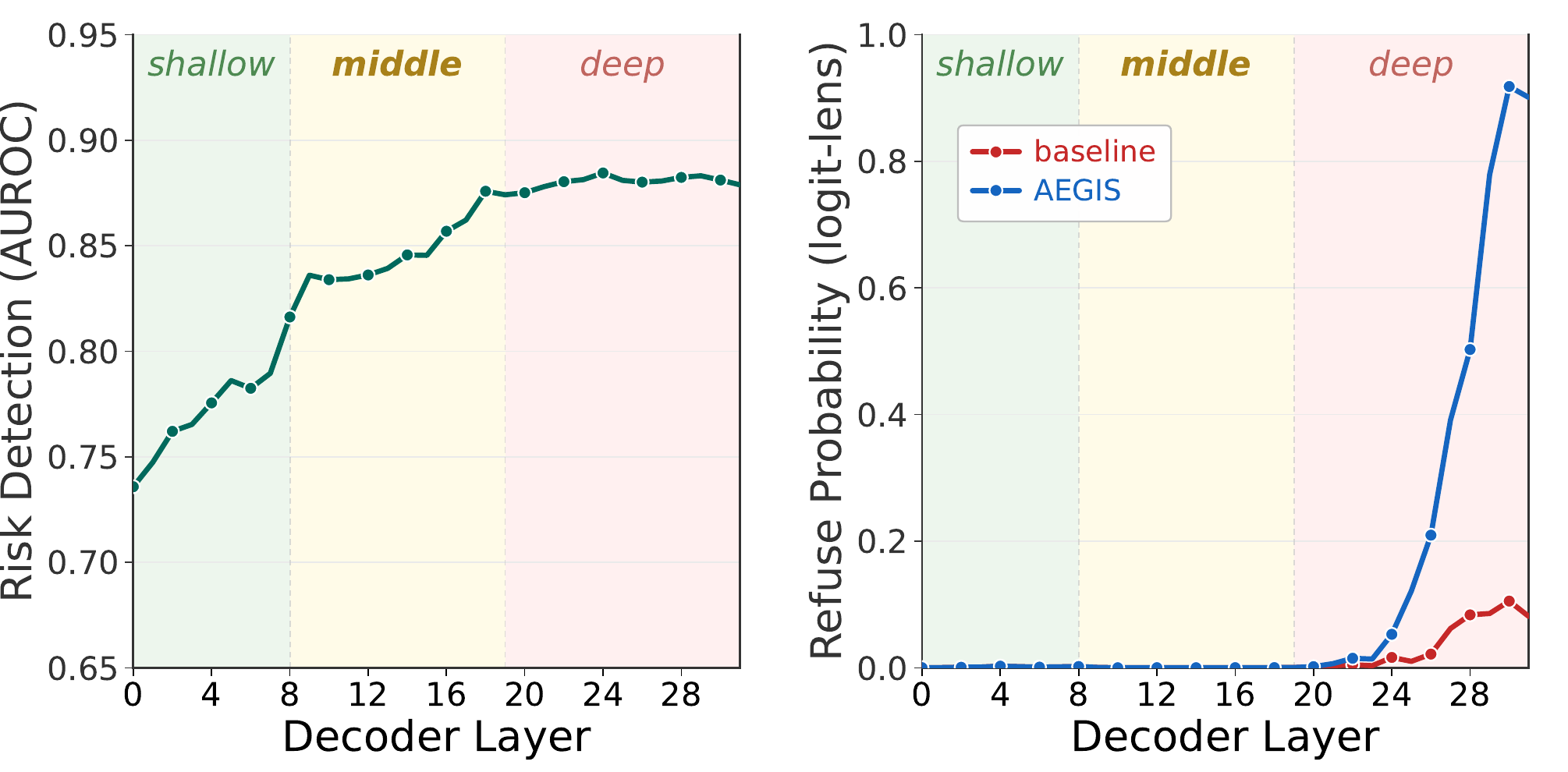}
    \caption{
    The \textit{risk-to-refusal gap} in Qwen2-Audio under successful audio jailbreaks.
    \textbf{Green} curves report layer-wise risk decodability (left), while \textbf{red} and \textbf{blue} curves report refusal-token probability before and after \textsc{Aegis} intervention, respectively (right). 
    \textsc{Aegis} raises late-layer refusal tendency, narrowing the gap.
    }
    \label{fig:observation}
\end{figure}

To answer this question, we trace two signals across the decoder layers of six undefended LALMs: risk decodability from internal representations and refusal tendency during generation. Fig.~\ref{fig:observation} illustrates the pattern using Qwen2-Audio as a representative example:  risk-related information (\textbf{green}) becomes increasingly decodable in middle and late layers, while refusal probability (\textbf{red}) remains near zero even in late layers. 
We observe a similar pattern across the other five models\footnote{Results for the remaining models are provided in the GitHub repository.} and call this discrepancy the \textit{risk-to-refusal gap}.

Building on this finding, we propose \textsc{Aegis} (in Fig.~\ref{fig:overview}), a detect-then-intervene defense that uses mid-layer risk information to selectively activate downstream safety intervention. A lightweight risk gate reads the hidden representation in a probe-selected mid-layer and produces a continuous risk score. This score selectively activates LoRA-based safety adapters in later layers, strengthening refusal for high-risk inputs while limiting intervention on benign inputs. 

We evaluate \textsc{Aegis} across six LALMs and three heterogeneous audio jailbreak benchmarks. Averaged across all evaluations under in-domain setting, \textsc{Aegis} reduces the unsafe rate from 17.9\% to 0.4\%, while marginally increasing over-refusal on benign inputs. Its performance under leave-one-benchmark-out evaluation further shows that the learned risk signal and intervention transfer across jailbreak distributions excluded from training. Beyond these behavioral gains, in Fig.~\ref{fig:observation}, \textsc{Aegis} also raises late-layer refusal tendency (\textbf{blue}), narrowing the \textit{risk-to-refusal gap}.

In summary, our contributions are: (1) We identify the risk-to-refusal gap between decodable risk-related information and weak refusal tendency, (2) we introduce \textsc{Aegis}, which converts retained risk information into selective late-layer safety intervention, and (3) across six LALMs and three heterogeneous jailbreak benchmarks, \textsc{Aegis} substantially reduces unsafe responses while maintaining limited over-refusal and effectiveness under held-out evaluation.

\section{Related Work}
Recent studies have revealed diverse safety vulnerabilities in audio-language models through broad benchmarks and red-teaming~\cite{peng2026jalmbenchbenchmarkingjailbreakvulnerabilities, song2025audiojailbreakopencomprehensive, yang2024audioachillesheelred, yang2025withstandchataudioattacksevaluation}. Existing attacks span semantic and compositional manipulations that reframe or distribute harmful intent~\cite{yu2026hearmeaudionarrative, yang2026speechaudiocompositionalattacksmultimodal}, acoustic and paralinguistic variations in language, accent, emotion, and speaking style~\cite{roh2025multilingualmultiaccentjailbreakingaudio, feng2026emotion, li2025stylebreakrevealingalignmentvulnerabilities}, and signal-level perturbations that optimize adversarial waveforms~\cite{peri2024speechguardexploringadversarialrobustness, kang2025advwave}. Their safety impact is commonly evaluated using harmful-behavior sets such as AdvBench~\cite{zou2023universaltransferableadversarialattacks} and JBB-Behaviors~\cite{chao2024jailbreakbenchopenrobustnessbenchmark}, while XSTest~\cite{röttger2024xstesttestsuiteidentifying} measures over-refusal on benign prompts. Together, these studies establish a heterogeneous audio safety landscape and motivate defenses that must remain robust across distinct attack mechanisms, while minimizing over-refusal on benign prompts.

Existing defenses intervene at different stages of the multimodal generation pipeline. Prompt-based methods strengthen safety instructions~\cite{peng2026jalmbenchbenchmarkingjailbreakvulnerabilities, roh2025multilingualmultiaccentjailbreakingaudio}, while guard-based approaches detect or filter unsafe inputs before generation~\cite{yang2026speechaudiocompositionalattacksmultimodal, verma2025omniguardefficientapproachai, ranjan2026voiceshieldsmallrealtimemaliciousspeech}. Other approaches modify model behavior through safety training~\cite{yang2025reshapingrepresentationspacebalance, wang2026omnisafetycrossmodalityconflictvulnerabilities, lu2025sea} or inference-time intervention~\cite{jin2025almguardsafetyshortcutsguardrails, djanibekov2025spiritpatchingspeechlanguage, lin2026sarsteersafeguardinglargeaudiolanguage}. Beyond defense, recent mechanistic studies show that harmfulness and refusal can be represented separately in LLMs and analyze how jailbreaks alter these internal safety signals~\cite{zhao2025harmfulnessrefusal}, and related multimodal work further studies layer-wise refusal dynamics under cross-modal safety conflicts~\cite{wang2026omnisafetycrossmodalityconflictvulnerabilities}. However, how retained risk information fails to translate into refusal under audio jailbreaks in LALMs remains less understood. We investigate this relationship through layer-wise probing and use the retained risk signal to guide selective safety intervention.

\section{Layer-wise Risk-to-Refusal Gap}

Successful audio jailbreaks reveal refusal failure, but not whether risk information is lost internally or remains encoded without triggering refusal. We distinguish these cases by tracing layer-wise risk decodability and refusal tendency.

\begin{figure}[t]
    \centering
    \includegraphics[width=\linewidth]{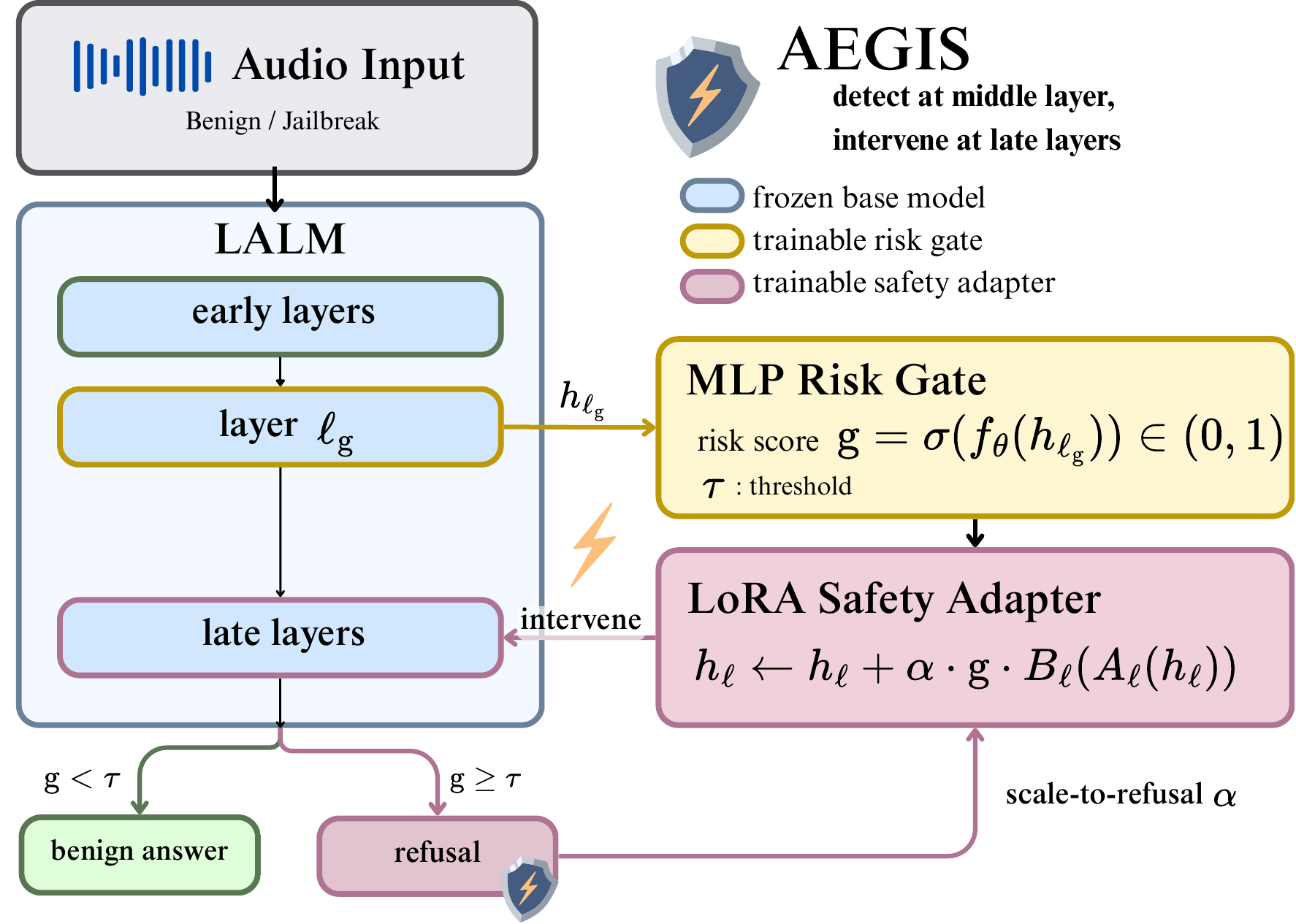}
    \caption{
    Overview of our proposed defense \textsc{Aegis}, a detect-then-intervene defense whose mid-layer risk gate selectively activates downstream safety adapters.
    }
    \label{fig:overview}
\end{figure}

\parhead{Layer-wise Risk Probing.}
We probe whether successfully jailbroken harmful inputs remain distinguishable from benign inputs in the model's internal representations. We construct a binary classification dataset of harmful inputs that elicit unsafe responses and benign inputs that elicit safe responses. For each input, we extract the final input token's hidden state at every decoder layer and train an independent linear probe, evaluated on a held-out split using AUROC. The resulting layer-wise AUROC measures how strongly harmful and benign inputs remain linearly separable, which we use as a proxy for the availability of risk-related information at each layer. 

\parhead{Layer-wise Refusal Tendency.}
We estimate refusal tendency using the unique first tokens of predefined refusal patterns as refusal-initiation tokens. At each generation step, we project each decoder layer's hidden state to the vocabulary space, apply softmax, and sum the probability mass assigned to these tokens. We then average this probability mass across generation steps to obtain a layer-wise refusal tendency score, which serves as a proxy for refusal tendency at each layer.

\begin{table*}[t]
\centering
\caption{
Unsafe rate (\%) of \textsc{Aegis} across six LALMs and three jailbreak benchmarks.
In-domain training includes the target benchmark, while LOBO training excludes it entirely.
Parentheses indicate changes from the undefended model.
\vspace{0.3em}}
\label{tab:overall_unsafe_rate}
\fontsize{9}{11}\selectfont
\setlength{\tabcolsep}{3pt}
\renewcommand{\arraystretch}{0.9}
\begin{tabular}{llllllll}
\toprule
Benchmark & Defense & Gemma 4 E4B & Phi-4 & ViTA-1.5 & Qwen2-Audio & Ultravox v0.5 & Voxtral Small \\
\midrule
\multirow{3}{*}{AJail (Origin) $\downarrow$}
 & No Defense & 7.90 & 4.40 & 6.40 & 12.55 & 7.20 & 18.90 \\
 & \textsc{Aegis} (in-domain) 
 & 0.30\,{\scriptsize(-7.60)}
 & 0.00\,{\scriptsize(-4.40)}
 & 0.00\,{\scriptsize(-6.40)}
 & 1.40\,{\scriptsize(-11.15)}
 & 0.40\,{\scriptsize(-6.80)}
 & 0.60\,{\scriptsize(-18.30)} \\
 & \textsc{Aegis} (LOBO)
 & 0.10\,{\scriptsize(-7.80)}
 & 0.00\,{\scriptsize(-4.40)}
 & 0.00\,{\scriptsize(-6.40)}
 & 0.80\,{\scriptsize(-11.75)}
 & 4.00\,{\scriptsize(-3.20)}
 & 10.30\,{\scriptsize(-8.60)} \\
\addlinespace[2pt]

\multirow{3}{*}{JALM (ADiv+SSJ) $\downarrow$}
 & No Defense & 16.90 & 66.80 & 1.60 & 16.70 & 4.40 & 22.80 \\
 & \textsc{Aegis} (in-domain)
 & 0.20\,{\scriptsize(-16.70)}
 & 1.20\,{\scriptsize(-65.60)}
 & 0.00\,{\scriptsize(-1.60)}
 & 0.00\,{\scriptsize(-16.70)}
 & 0.00\,{\scriptsize(-4.40)}
 & 0.60\,{\scriptsize(-22.20)} \\
 & \textsc{Aegis} (LOBO)
 & 7.50\,{\scriptsize(-9.40)}
 & 7.60\,{\scriptsize(-59.20)}
 & 0.00\,{\scriptsize(-1.60)}
 & 0.80\,{\scriptsize(-15.90)}
 & 2.40\,{\scriptsize(-2.00)}
 & 13.20\,{\scriptsize(-9.60)} \\
\addlinespace[2pt]

\multirow{3}{*}{SACRED (MSD) $\downarrow$}
 & No Defense & 17.20 & 46.40 & 2.40 & 42.45 & 0.40 & 26.40 \\
 & \textsc{Aegis} (in-domain)
 & 0.00\,{\scriptsize(-17.20)}
 & 1.30\,{\scriptsize(-45.10)}
 & 0.00\,{\scriptsize(-2.40)}
 & 1.60\,{\scriptsize(-40.85)}
 & 0.00\,{\scriptsize(-0.40)}
 & 0.00\,{\scriptsize(-26.40)} \\
 & \textsc{Aegis} (LOBO)
 & 0.00\,{\scriptsize(-17.20)}
 & 0.00\,{\scriptsize(-46.40)}
 & 0.00\,{\scriptsize(-2.40)}
 & 0.00\,{\scriptsize(-42.45)}
 & 0.00\,{\scriptsize(-0.40)}
 & 0.40\,{\scriptsize(-26.00)} \\
\midrule

Over-Refusal (\%) $\downarrow$
& No Defense
& 10.8 & 11.6 & 12.0 & 36.80 & 26.8 & 3.2 \\
& \textsc{Aegis} (in-domain)
& +10.4 & +2.4 & +0.5 & +9.3 & +7.5 & +7.7 \\
& \textsc{Aegis} (LOBO)
& +10.1 & +9.9 & +5.1 & +5.3 & +1.3 & +9.6 \\

\bottomrule
\end{tabular}
\end{table*}
\parhead{Observation.}
Fig.~\ref{fig:observation} reveals a consistent dissociation between the two signals under the undefended model. The \textbf{green} curve shows that risk-related information becomes increasingly decodable in middle and deep layers, even for inputs that ultimately elicit unsafe responses, whereas the \textbf{red} curve shows that refusal tendency remains near zero. This pattern suggests that risk-related information remains decodable without a corresponding increase in refusal tendency. We refer to this mismatch as the \textit{risk-to-refusal gap}, which motivates \textsc{Aegis} to detect the retained mid-layer risk signal and selectively intervene in later layers.

\section{AEGIS: Detect-then-Intervene Defense}
Building on the observation of the \textit{risk-to-refusal gap}, we propose \textsc{Aegis}, a defense using internal risk signals from mid-layer representations to control downstream safety intervention. The base LALM remains frozen while two lightweight components are jointly trained: (1) a risk gate attached to a mid-layer and (2) LoRA safety adapters inserted into the late layers. Fig.~\ref{fig:overview} provides an overview.

\parhead{Mid-layer MLP Risk Gate.}
Given an audio input $x_a$, \textsc{Aegis} 
extracts the final prompt-token representation $h_{\ell_g}$ from an mid-layer $\ell_g$, selected separately for each model by layer-wise probing analysis. A lightweight multilayer perceptron (MLP) maps this representation to a continuous risk score:
\[
g = \sigma\left(f_\theta(h_{\ell_g})\right),
\]
where $f_\theta$ denotes the MLP, $\sigma$ is the sigmoid function, and $g \in (0,1)$ estimates the risk that the input is unsafe and controls the downstream safety intervention.

\parhead{Late-layer Safety Adapters.}
Using the gate score $g$, we condition LoRA-based safety adapters inserted into a set of later layers $\mathcal{L}_{\mathrm{int}}$. When the intervention is activated, the hidden state $h_\ell$ in each layer $\ell\in \mathcal{L}_{\mathrm{int}}$ is updated as 
\[
h_\ell \leftarrow h_\ell + \alpha \cdot g \cdot B_\ell(A_\ell(h_\ell)),
\]
where $A_\ell$ and $B_\ell$ are low-rank trainable projection matrices, $g$ is the gate-predicted risk score and $\alpha$ controls the intervention strength. 
Multiplying the low-rank residual update by $g$ strengthens the safety intervention for high-risk inputs while minimizing its effect on low-risk inputs. At inference, only inputs above the risk threshold activate the adapters. For activated inputs, $\alpha$ progressively increases through closed-loop scaling until a predefined refusal criterion is met.

\parhead{Training Objective.}
We jointly train the risk gate and low-rank safety adapters with the frozen base model. The training objective is 
\[
\mathcal{L}
=
\mathcal{L}_{\mathrm{LM}}
+
\lambda_{\mathrm{BCE}} \cdot \mathrm{BCE}(g, z)
+
\lambda_{\mathrm{L1}} \cdot g,
\]
where $z$ is the binary safety label, with $z=1$ for harmful examples and $z=0$ for benign examples.
The loss $\mathcal{L}_{\mathrm{LM}}$ is computed only on the target response tokens, with the prompt tokens masked. The binary cross-entropy term $\lambda_{\mathrm{BCE}}=1.0$ supervises risk estimation, while the sparsity penalty $\lambda_{\mathrm{L1}}=0.01$ discourages unnecessary gate activation. Together, these objectives encourage \textsc{Aegis} to detect and intervene on harmful inputs while minimizing changes to benign behavior.

\section{Experiments}

\subsection{Experimental Setup}

\parhead{Models.} We evaluate on six LALMs from different model families: Qwen2-Audio-7B~\cite{Qwen2Audio}, VITA-1.5~\cite{fu2025vita}, Ultravox v0.5~\cite{ultravox}, Phi-4-Multimodal Instruct~\cite{microsoft2025phi4minitechnicalreportcompact}, Voxtral-Small-24B~\cite{liu2025voxtral}, and Gemma 4 E4B Instruct~\cite{gemmateam2026gemma4}.

\parhead{Datasets.}
We use three audio jailbreak benchmarks: the AJailBench~\cite{song2025audiojailbreakopencomprehensive} \emph{Origin} subset (1{,}490 samples), the JALMBench~\cite{peng2026jalmbenchbenchmarkingjailbreakvulnerabilities} SSJ and ADiv subsets (246 and 700 samples), and the SACRED-Bench~\cite{yang2026speechaudiocompositionalattacksmultimodal} MSD subset (1{,}364 samples), covering signal-level perturbations, semantic obfuscation, multilingual attacks, and compositional jailbreaks in multi-speaker dialogue. 
For benign training, we use TTS-converted benign prompts from JBB-Behavior~\cite{chao2024jailbreakbenchopenrobustnessbenchmark} and LLM-generated general benign instructions.
For benign evaluation, we synthesize an audio version of XSTest~\cite{röttger2024xstesttestsuiteidentifying} using Qwen3-TTS~\cite{Qwen3-TTS}.

\parhead{Evaluation Protocol.}
We evaluate \textsc{Aegis} under both in-domain and leave-one-benchmark-out (LOBO) settings. In the in-domain setting, training includes training samples from the benchmark being evaluated, while evaluation is conducted on its held-out split. For LOBO evaluation, we hold out one jailbreak benchmark entirely and train \textsc{Aegis} on the remaining two benchmarks before evaluating on the held-out benchmark.

\parhead{Metrics.}
Unsafe rate is the percentage of model outputs classified as unsafe by Llama-Guard-3-8B~\cite{dubey2024llama3herdmodels}.
Over-refusal rate is the percentage of responses to benign XSTest inputs that match a predefined refusal-pattern regular expression. Across all tables, jailbreak results report unsafe rate (\%), and OR denotes absolute over-refusal rate, while $\Delta$OR denotes its change from the undefended model in percentage points.

\parhead{Baselines.}
We compare \textsc{Aegis} (LOBO) against training-based defenses RRS~\cite{yang2025reshapingrepresentationspacebalance} and OmniSteer~\cite{wang2026omnisafetycrossmodalityconflictvulnerabilities}, as well as training-free ALMGuard~\cite{jin2025almguardsafetyshortcutsguardrails} and SARSteer~\cite{lin2026sarsteersafeguardinglargeaudiolanguage}. We retrain these baselines on the same data as \textsc{Aegis}.
We also evaluate an ASR-based preprocessing baseline using Whisper-large-v3~\cite{radford2022whisper}. 
As a prompt-based baseline (PBD), we prepend the defensive system prompt from JailbreakBench (JBB) ~\cite{yang2026speechaudiocompositionalattacksmultimodal} to the target model's system prompt.
Because several baselines require model-specific internal access or additional training, we conduct all defense comparisons on Qwen2-Audio under the same evaluation benchmarks and metrics.

\subsection{Overall Safety, Utility, and Design Effect}

\begin{table}[t]
\centering
\caption{
Ablation study of \textsc{Aegis} on Qwen2-Audio under LOBO evaluation. 
\vspace{0.3em}}
\label{tab:qwen_ablation}
\fontsize{9}{11}\selectfont
\setlength{\tabcolsep}{3pt}
\renewcommand{\arraystretch}{0.9}

\begin{tabular}{lcccc}
\toprule
\textbf{Variant}
&
\textbf{AJail} $\downarrow$
&
\textbf{JALM} $\downarrow$
&
\textbf{SACRED} $\downarrow$
&
\textbf{$\Delta$OR} $\downarrow$
\\
\midrule

Full
& \textbf{0.3}
& \textbf{0.0}
& 2.0
& +22.1 \\

Always-on
& \textbf{0.3}
& 2.0
& 2.2
& +21.9 \\

\textsc{Aegis} (LOBO)
& 0.8
& 0.8
& \textbf{0.0}
& \textbf{+5.30} \\

\bottomrule
\end{tabular}
\end{table}

\begin{table}[t]
\centering
\caption{
Comparison with representative defenses on Qwen2-Audio.
\bestshade{Darker shading} and \secondshade{lighter shading}
indicate the best and second-best performance among defense methods
in each column, respectively.
}
\label{tab:baseline_comparison}

\fontsize{9}{11}\selectfont
\setlength{\tabcolsep}{2pt}
\renewcommand{\arraystretch}{0.9}
\begin{tabular}{lrrrr}

\toprule

\textbf{Method}
& \textbf{AJail} $\downarrow$
& \textbf{JALM} $\downarrow$
& \textbf{SACRED} $\downarrow$
& $\Delta$\textbf{OR} $\downarrow$\\

\midrule

Vanilla
& \basecell{12.55}
& \basecell{16.70}
& \basecell{42.45}
& \basecell{0.00} \\

\midrule

RRS (EMNLP'25)
& 11.61
& 26.32
& 38.71
& +11.33 \\

OmniSteer (arXiv'26)
& 12.21
& 17.02
& 42.96
& \bestcell{+0.80} \\

ALMGuard (NeurIPS'25)
& 13.09
& 22.73
& 43.55
& +12.00 \\

SARSteer (ICML'26)
& \secondcell{4.43}
& \secondcell{6.13}
& 26.91
& +41.60 \\

ASR (OpenAI'23)
& 12.15
& 31.09
& \secondcell{3.17}
& \secondcell{+1.40} \\

PBD-JBB (NeurIPS'24)
& 9.73
& 26.96
& 33.94
& +12.00 \\

\midrule

\textsc{Aegis} (Ours, LOBO)
& \bestcell{0.80}
& \bestcell{0.80}
& \bestcell{0.00}
& +5.30 \\

\bottomrule

\end{tabular}
\end{table}

Table~\ref{tab:overall_unsafe_rate} shows that \textsc{Aegis} remains effective despite substantial variation in baseline vulnerability across models and benchmarks. Across the 18 model-benchmark pairs, \textsc{Aegis} reduces the average unsafe rate from 17.9\% to 0.4\% under in-domain training. This improvement indicates that the effectiveness of \textsc{Aegis} is not tied to a particular model family or safety profile. The safety gain also comes with a limited utility cost, marginally increasing XSTest over-refusal by 6.3 percentage points on average. 
To assess whether these gains depend on exposure to the evaluated benchmark during training, we also examine the LOBO results. \textsc{Aegis} remains effective when the target benchmark is entirely excluded from training, reducing the average unsafe rate to 2.6\%.

Table~\ref{tab:qwen_ablation} further clarifies the role of selective intervention using two variants. \textit{Full} applies safety LoRA across all decoder layers, while \textit{Always-on} retains the late-layer adapters but removes the risk gate, intervening on every input.
Both variants achieve strong protection but increase XSTest over-refusal by 22.1 and 21.9 percentage points, compared with 5.3 points for \textsc{Aegis}. This indicates that the risk gate mainly controls when intervention is applied, preserving safety while avoiding unnecessary refusal on benign inputs.

\subsection{Comparison with Representative Defenses}
Table~\ref{tab:baseline_comparison}  compares \textsc{Aegis} with representative defenses under the same Qwen2-Audio evaluation setting. \textsc{Aegis} is the only method that maintains an unsafe rate below 1\% across all three benchmarks, whereas existing defenses often show strong gains on only a subset of attacks. For example, SARSteer substantially improves AJail and JALM, but remains at 26.9\% on SACRED.
This consistency is also achieved without excessive over-refusal. \textsc{Aegis} increases XSTest over-refusal by only 5.3 percentage points, compared with substantially larger increases for most competing defenses. 
Overall, \textsc{Aegis} achieves a favorable safety-utility trade-off by maintaining strong protection across heterogeneous attacks while limiting unnecessary refusal on benign inputs.

\subsection{Closing the Risk-to-Refusal Gap}

Beyond improving output-level safety, \textsc{Aegis} also alters the internal dynamics underlying successful jailbreaks. As shown in Fig.~\ref{fig:observation}, the late-layer refusal tendency (\textbf{blue}) rises sharply after intervention, in contrast to the near-zero tendency (\textbf{red}). This indicates that \textsc{Aegis} 
reconnects decodable risk-related information with late-layer refusal tendency, thereby substantially narrowing the \textit{risk-to-refusal gap}.

\section{Conclusion}
In this work, we identify the \textit{risk-to-refusal gap}, where risk-related information remains internally decodable while refusal tendency remains weak under successful audio jailbreaks. Building on this observation, we introduce \textsc{Aegis}, a detect-then-intervene defense that selectively converts retained mid-layer risk signals into late-layer safety interventions. Across six LALMs and heterogeneous jailbreak benchmarks, \textsc{Aegis} consistently improves safety under in-domain and LOBO settings, and narrows the gap while limiting over-refusal.

\parhead{Limitations.}
Our analysis relies on proxy measures that may not fully capture internal safety mechanisms. \textsc{Aegis} requires access to model internals, limiting its applicability to closed-source LALMs. Finally, while our evaluation covers diverse attack mechanisms, it does not encompass the full range of audio jailbreaks or adaptive attacks targeting the defense itself.

\section{Compliance with Ethical Standards}
This study uses existing public benchmarks and collects no new human-subject data. No ethical approval was required.

\bibliographystyle{IEEEbib}
\bibliography{main}

\end{document}